\documentclass[
    reprint,                
    superscriptaddress,     
    amsmath,                
    amssymb,                 
    aps,                    
    prl,                    
    floatfix                 
]{revtex4-1}
\usepackage{siunitx}

\usepackage{graphicx}       
\usepackage[dvipsnames]{xcolor} 
\usepackage{bm}             
\usepackage{amsbsy}         
\usepackage{float}          

\usepackage[T1]{fontenc}    
\usepackage[utf8]{inputenc} 
\usepackage{lmodern} 
\usepackage{amssymb}

\allowdisplaybreaks

\usepackage[
    colorlinks=true,
    linkcolor=MidnightBlue,
    citecolor=OliveGreen,
    urlcolor=BrickRed,
    pdfauthor={Your Name},
    pdftitle={Your Paper Title},
    pdfsubject={},
    pdfkeywords={}
]{hyperref}

\begin{document}

\title{The Integration Host Factor is a pH-responsive protein that switches from DNA bending to DNA bridging in acidic biofilm-like conditions}

\author{Dinesh Parthasarathy}
\thanks{joint first author}
\affiliation{School of Physics and Astronomy, University of Edinburgh, Peter Guthrie Tait Road, Edinburgh EH9 3FD, United Kingdom}
\author{Saminathan Ramakrishnan}
\thanks{joint first author}
\affiliation{School of Physics and Astronomy, University of Edinburgh, Peter Guthrie Tait Road, Edinburgh EH9 3FD, United Kingdom}
\author{Georgia Tsang}
\affiliation{School of Biological Sciences, University of Edinburgh, Colin Maclaurin Road, Edinburgh EH9 3DW, United Kingdom}
\author{Auro Varat Patnaik}
\affiliation{School of Physics and Astronomy, University of Edinburgh, Peter Guthrie Tait Road, Edinburgh EH9 3FD, United Kingdom}
\author{Sabrina M.C. Hardy}
\affiliation{School of Physics and Astronomy, University of Edinburgh, Peter Guthrie Tait Road, Edinburgh EH9 3FD, United Kingdom}
\author{Willem Vanderlinden}
\affiliation{School of Physics and Astronomy, University of Edinburgh, Peter Guthrie Tait Road, Edinburgh EH9 3FD, United Kingdom}
\author{Jamieson Howard}
\affiliation{School of Physics, Engineering \& technology, University of York, YO10 5DD, United Kingdom}
\author{Braden Bylett}
\affiliation{School of Physics, Engineering \& technology, University of York, YO10 5DD, United Kingdom}
\author{James R. Law}
\affiliation{School of Physics, Engineering \& technology, University of York, YO10 5DD, United Kingdom}
\author{Mark C. Leake}
\affiliation{School of Physics, Engineering \& technology, University of York, YO10 5DD, United Kingdom}
\affiliation{Department of Biology, University of York, YO10 5DD, United Kingdom}
\author{Agnes Noy}
\thanks{corresponding author}
\affiliation{School of Physics, Engineering \& technology, University of York, YO10 5DD, United Kingdom}
\author{Davide Michieletto}
\thanks{corresponding author}
\affiliation{School of Physics and Astronomy, University of Edinburgh, Peter Guthrie Tait Road, Edinburgh EH9 3FD, United Kingdom}
\affiliation{MRC Human Genetics Unit, Institute of Genetics and Cancer, University of Edinburgh, Edinburgh EH4 2XU, United Kingdom}

\newcommand{\dmi}[1]{\textcolor{RoyalBlue}{#1}}
\newcommand{\din}[1]{\textcolor{Orange}{#1}}

\begin{abstract}
The Integration Host Factor (IHF) is a nucleoid-associated protein critical for both DNA compaction and biofilm stability. While its role in DNA packaging within the cell is well understood, its structural role in scaffolding biofilms is more puzzling and difficult to reconcile with its known DNA bending activity. Here, we investigated how IHF-DNA interactions are modulated across a pH spectrum mimicking the acidic microenvironments of bacterial biofilms. By performing all-atom calculations we discovered that low pHs lead to a change in protonation of IHF residues, which in turn exposes positively charged patches. We then conjectured that these positively charged residues could lead to intermolecular DNA bridging and tested this hypothesis through single-molecule and bulk assays. We discovered that while at physiological pH IHF mostly bends DNA, at pH $<$ 5 there is clear evidence of IHF-mediated intermolecular crosslinking. Our results demonstrate that pH significantly modulates IHF-DNA interactions and explains the structural role played by IHF in supporting biofilm mechanics through intermolecular crosslinking. 
\end{abstract}

\maketitle

\section{Introduction}
\begin{figure}[t!]
    \centering
    \includegraphics[width=0.5\textwidth]{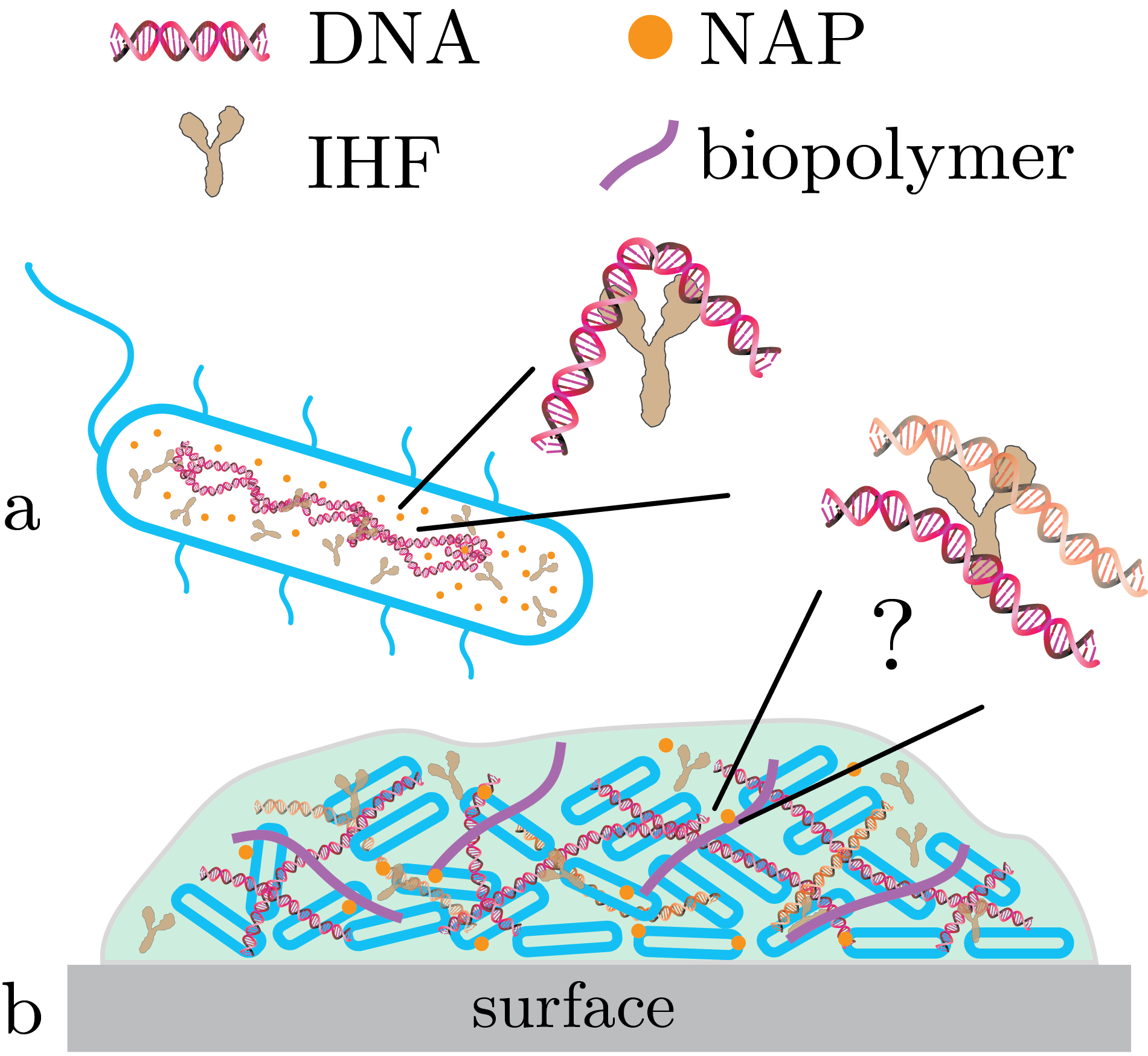}
    \caption{\textbf{Roles of IHF in cells and biofilms.} \textbf{a} Within bacterial cells, IHF contributes to genome packaging by inducing sharp DNA bending~\cite{Yoshua2021IHF}. \textbf{b} In the biofilm, IHF contributes to strengthening the extracellular matrix~\cite{novotny2013structural}. We lack a quantitative understanding of the mechanisms underpinning the role of IHF in the mechanical stability of biofilms.}
    \label{fig:dual_role_IHF}
\end{figure}

Across all domains of life, DNA is densely packed in cells that are between 1,000 and 100,000 times smaller than their genome~\cite{alberts2002cell}. To achieve this level of packaging, bacteria and eukaryotes have evolved a range of DNA binding proteins that guide DNA organisation, e.g. nucleoid associated proteins (NAPs) and histones, respectively~\cite{alberts2002cell,dame2005role,schwab2025histone}. One of the most common NAP in bacteria is the Integration Host Factor (IHF), a heterodimeric NAP with a well-characterized DNA-binding activity~\cite{rice1996crystal, murphy1997isolation, Yoshua2021IHF}. IHF introduces sharp bends in DNA which are key for organizing and compacting the bacterial genome within the nucleoid~\cite{dame2005role} (Fig.~\ref{fig:dual_role_IHF}). 

The molecular basis of IHF-mediated DNA bending has been elucidated through structural analyses~\cite{rice1996crystal}. The protein core contains entwined $\alpha$-helices from each subunit, with extended $\beta$-ribbon arms that penetrate the minor groove upon binding~\cite{rice1996crystal}. Each arm terminates with a conserved proline residue, which intercalates between base pairs, disrupting stacking interactions and introducing two kinks $\sim$ 9 bp apart. This arrangement facilitates DNA bending angles exceeding 160$^{o}$~\cite{Yoshua2021IHF}, while the positively charged protein surface stabilizes the protein-DNA complex~\cite{rice1996crystal}.

At the same time, while bacteria can exist as isolated cells in cell cultures, in natural settings most thrive in multicellular communities called biofilms ~\cite{flemming2010biofilm} which are structurally shaped by the bacteria themselves through the secretion of nucleic acids, biopolymers and proteins~\cite{Seifert2017,Jurcisek2017} and through programmed cell death or cell lysis~\cite{Turnbull2016Explosive,Prentice2024CellLysis}. NAPs mechanistically contribute to supporting biofilm integrity by interacting with extracellular and environmental genomic DNA (eDNA)~\cite{goodman2011biofilms}. 
Using anti-IHF antibodies, it was shown that IHF contributes to the structural integrity of the presence of \textit{B. cenocepacia} biofilms \textit{in vitro} and confirmed the presence of IHF within a sputum sample from a Cystic Fibrosis (CF) patient~\cite{novotny2013structural}. Pre-treatment of biofilms using anti-IHF showed significant reductions in biofilm integrity: 44$\%$ reduction in thickness, 56 $\%$ reduction in biomass and 52 $\%$ reduction in height, in comparison to pre-treatment with naive serum~\cite{novotny2013structural}.
More generally, inhibiting NAPs and DNaseI treatment both result in biofilm destabilisation and heightened bacterial susceptibility to antimicrobials~\cite{goodman2011biofilms,Kaplan2012Recombinant}. Thus, NAPs - and IHF in particular -- are critical biofilm components and provide architectural support to the biofilm’s matrix~\cite{payne2016emerging,Devaraj2019,Sharma2024eDNAbiofilm} (Fig.~\ref{fig:dual_role_IHF}). However, the mechanism through which IHF reinforces biofilm stability is not fully understood. If IHF were bending eDNA in the biofilm it would be expected to decrease entanglements and thus fluidify the biofilm rather than reinforce it~\cite{fosado2023fluidification}. We thus currently lack a model to explain how IHF interacts with eDNA (and potentially other polysaccharides) in the biofilm in order to provide structural support to the matrix.

Biofilms of various bacteria, including the common \textit{Pseudomonas aeruginosa} biofilms display pH values as low as 5~\cite{wilton2016extracellular,ChokyuDelRey2024}, becoming more acidic the deeper into the biofilm matrix~\cite{kromer2022monitoring,Hou2017pHbiofilm,Xiao2017Biofilm}. On contrary, in the cytoplasm IHF is exposed to pH neutral conditions. We thus specifically asked whether acidic conditions in the biofilm affect IHF interaction with (e)DNA. Motivated by this, in this work we systematically investigated the interaction of IHF with DNA in acidic conditions using both computational and experimental methods. 

We employed all atom simulations, single molecule (optical tweezers) and bulk (microrheology) assays to quantify non-specific interactions of IHF with DNA at different pH conditions (7.5, 6.5, 5.5 and 4.5) to mimick the varying acidic conditions in biofilms. Atomistic calculations suggest that surface-exposed IHF residues become protonated at low pH, potentially presenting sticky sites for the negatively charged DNA. Our single molecule and bulk assays support this model, and they suggest that IHF switches from DNA-bending to DNA-bridging at pH $\simeq$ 5.5, in turn acting as a non-specific crosslinker of negatively charged polymers, such as eDNA. 

\section{Materials and methods}

\subsection{IHF expression and purification}

IHF was produced in the \textit{E. coli} strain BL21AI containing the plasmid pRC188 (a gift from the Chalmers laboratory, the University of Nottingham, UK). The \emph{E. coli} harbouring pRC188 were grown to an OD600 $\sim$ 0.6 in 2L LB + 100 $\mu$g/ml carbenicillin at 37$^\circ$C with shaking at 180 rpm. Induction of IHF overexpression was carried out by the addition of arabinose and IPTG to respective final concentrations  of 0.2\% (w/v) and 1 mM, growth was then allowed to continue for a further 3 hours at 37$^\circ$C with shaking at 180 rpm. The \emph{E. coli} were collected via centrifugation at 3500 x g for 20 minutes at 4$^\circ$C. After discarding the supernatant cells were resuspended in 10 mM Tris pH 7.5, 10\% sucrose (w/v) before being flash frozen in liquid nitrogen and stored at -80$^\circ$C.
Cells overexpressing IHF were thawed on ice and the buffer was adjusted to final concentrations of 50 mM Tris pH 8.4, 150 mM KCl, 20 mM EDTA, 10 mM DTT and 0.2 mg/ml lysozyme. The cell suspension was mixed by inversion and kept on ice for 15 minutes before Brij58 (Sigma-Aldrich) was added to a final concentration of 0.1\% (w/v) followed by further mixing by inversion and a 15 minute incubation on ice. The lysed cells were clarified by centrifugation at 4$^\circ$C and 148,000 x g for 60 minutes. The supernatant was collected and polymin P was added to a final concentration of 0.075\% (w/v) in a dropwise fashion whilst stirring at 4$^\circ$C, stirring was continued for 10 mins before centrifugation at 4$^\circ$C and 30,600 x g for 20 minutes. The resulting supernatant was then subjected to a 50\% ammonium sulfate (AmSO$_{4}$) precipitation followed by an 80\% AmSO$_{4}$ precipitation. Pellets and supernatant were collected following centrifugation as above with polymin P, IHF was present in the supernatant at 50\% AmSO$_{4}$ and in the pellet at 80\% AmSO$_{4}$. AmSO$_{4}$ precipitated IHF was dissolved in a sufficient volume of buffer A (50 mM Tris-HCl pH 7.5, 2 mM EDTA, 10 mM $\beta$-ME, 10\% glycerol) such that the conductivity of the sample matched that of buffer A + 100 mM KCl. The sample was loaded onto a self-poured  10 mL P-11 phosphocellulose XK-26 column equilibrated with buffer A + 100 mM KCl, washed with 30 column volumes (CV) of buffer A + 100 mM KCl and developed with a 20 CV gradient of 0.1-1 M KCl in buffer A. Fractions containing IHF were identified by their absorbance at the wavelength of 280 nm and with 15\% SDS polyacrylamide gel electrophoresis, these fractions were pooled. Pooled fractions were dialyzed against buffer A + 100 mM NaCl. The sample was then loaded onto a 5 mL HiTrap Heparin column equilibrated with the same buffer, washed with 6 CV of buffer A + 100 mM NaCl and  developed with a 20 CV gradient of 0.1-1 M NaCl in buffer A. Fractions containing IHF were again identified using absorbance at 280 nm  and 15\% SDS polyacrylamide gel electrophoresis and pooled. Pooled fractions were dialysed against buffer B (25 mM Tris.HCl pH 7.5, 550 mM KCl and 40\% glycerol), the dialysed fractions were then aliquoted and flash frozen in liquid nitrogen before storage at -80$^\circ$C. Protein concentrations were determined using the Bradford Protein Assay (Bio-Rad).

\subsection{Optical tweezer measurements}

The force-extension curves of DNA-IHF complexes across various pH conditions were recorded using a LUMICKS C-trap dual-trap optical tweezer (OT) system. Prior to experiments, each microfluidic channel was thoroughly washed with HPLC-grade water and 1X PBS. To further minimize non-specific binding, the protein channel was passivated with BSA (0.1 $\%$ w/v in PBS). Streptavidin-coated beads (1:100 dilution in PBS), biotinylated $\lambda$ DNA (1:1000 dilution), and IHF protein (150~nM) in IHF buffer (20 mM Tris, 50 mM KCl, 2 mM $\mathrm{MgCl_2}$, 7$\%$ glycerol) were introduced into their respective channels for a period of 10 minutes. Streptavidin-coated beads ( 4.0-4.9 $\mu$m, 0.5$\%$ (w/v)) and biotinylated bacteriophage $\lambda$-DNA (20ng/$\mu$L) were purchased from LUMICKS. Beads were first optically trapped, and were subsequently transferred to the buffer channel for optical trap stiffness calibration using C-trap bluelake software (V2.0). Subsequently, the beads were moved to the DNA channel to selectively capture and tether individual DNA molecules. Flow stretching ensured that the biotinylated ends of $\lambda$ DNA were securely attached to the streptavidin-coated beads. For each experiment, single-DNA tethering was confirmed by recording and verifying the characteristic force-extension curve of $\lambda$ DNA stretched between $\sim$ 1-55 pN. Force-extension (F-d) curves were initially collected in an IHF-free buffer. The DNA was then transferred into an IHF-containing buffer, incubated for 1 minute, and the Force-extension curve was measured at constant speed of 500 nm/s with at least five continuous measurements for each condition.

\subsection{Optical tweezer data analysis}
The F-d curves obtained from OT can be fitted, in the absence of proteins, via a semi-flexible worm-like chain (WLC) model~\cite{manning2006persistence}. However, when proteins, e.g. IHF, bind to DNA, they can alter both its bending and twisting stiffness. To capture these effects, we used the twistable worm-like chain (TWLC) model~\cite{gross2011quantifying}, which incorporates both bending (persistence length) and twisting elasticity (twist persistence length). The TWLC model relates the extension $x$ of a DNA molecule under an applied force $F$ to its contour length $L_c$ as:
\begin{equation}
x = L_c \left( 1 - \frac{1}{2} \sqrt{\frac{k_B T}{F \cdot L_p}}
+ \frac{C}{-g(F)^2 + S C} \cdot F \right),
\label{eq:twlc}
\end{equation}
where $k_B$ is the Boltzmann constant, $T$ is the temperature, $L_p$ is the bending persistence length, $C$ is the twist persistence length,  $S$ is the stretch modulus, and $g(F)$ is the force-dependent twist–stretch coupling term.
Fitting of the experimental force-extension (F-d) curves was performed using the \text{lumicks.pylake} analysis package, which provides an implementation of the twistable worm-like chain (TWLC) model. The fitting was carried out in two stages to improve parameter convergence. First, the low-force regime of the data (below $\sim$30 pN) was fitted using the extensible Odijk worm-like chain model \cite{odijk1995stiff}, yielding initial estimates for the contour length $L_c$ and persistence length $L_p$. These values were then used as starting parameters for the TWLC fit.
Literature values were used to fix the twist rigidity $C = 440$ pN/nm$^2$ and the force $F_c = 30.6$ pN~\cite{broekmans2016dna}, while $L_c$ and $L_p$ were treated as free parameters in the fit. The final fit was performed over the full force range (0-60 pN), yielding the extension as a function of applied force according to Equation \ref{eq:twlc}. This was done with a custom made script, deposited at \url{https://git.ecdf.ed.ac.uk/taplab/twistedWorm/}. 


\subsection{AFM sample preparation and image analysis}
To visualize IHF-DNA complexes under AFM at a pH range from 7.5 to 5.5 we employed linearized pUC19 (2.686 kbp). In IHF buffer, 1 nM of pUC19 was mixed with 150 nM of IHF and incubated at room temperature for 15 minutes. IHF buffers with specific pH were used for the respective conditions. 5 $\mu$L of the IHF-DNA complex sample was deposited on a Poly-L-Lysine pretreated mica sample for 1 minute, gently washed with 20 ml of deionized water, and air dried\cite{vanderlinden2014structure}. For the control DNA sample, only pUC19 DNA was mixed in the IHF buffer and prepared the same way. The AFM images of the DNA and IHF-DNA complexes were recorded using a Bruker JPK NanoWizard 4XP instrument with FASTSCAN-A probes.

AFM topographs were exported from the microscope software in ASCII format and processed using a custom Python pipeline \cite{kolbeck2024hiv}. Each data file contained pixel height values (z-coordinates) on a regularly spaced grid with defined lateral dimensions; pixel size (in nm) was calculated by dividing the scan size by the number of pixels per side.

Images were preprocessed by applying two stages of background thresholding to minimize noise and eliminate substrate signal. The initial threshold, applied after Gaussian blurring, removed most of the flat substrate and low-lying noise. Subsequent Gaussian filtering and a second threshold refined molecular boundaries and suppressed residual background. The molecules were then segmented by connected-component labeling of thresholded images, and objects below a minimum area were excluded. Skeletons of segmented molecules were analyzed for endpoints, branchpoints, and local pixel height to permit morphological categorization. Only bare DNA molecules and DNA-protein complexes were retained for analysis, while molecules not meeting these criteria were discarded.

For each retained molecule, the radius of gyration ($R_g$) was calculated to quantify structural compactness. Pixel coordinates were converted to nanometres using the calibrated pixel size. The intensity-weighted center of mass ($\mathbf{r}_{\text{cm}}$) was computed from the pixel positions $\mathbf{r}_i$ and their corresponding AFM height values $I_i$. The mass-weighted radius of gyration was then obtained as
\[
R_g^2 = \frac{\sum_i I_i \, \lVert \mathbf{r}_i - \mathbf{r}_{\text{cm}} \rVert^2}{\sum_i I_i},
\]
with 
\[
\mathbf{r}_{\text{cm}} = \frac{\sum_i I_i \, \mathbf{r}_i}{\sum_i I_i}.
\]
where $I_i$ denotes the pixel height at coordinate $\mathbf{r}_i$. This formulation accounts for the spatial distribution of the height signal within each molecule, rather than treating all pixels equally. All segmentation and classification results were manually verified. The molecule-level parameters (including $R_g$) were exported for statistical analysis and the representative close-up images of each molecule were also generated for visual inspection, retaining only isolated single DNA molecules while discarding aggregates, overlaps, and artifacts.

\subsection{Preparing dense $\lambda$DNA solutions and their controlled acidification}
Approximately 300 $\mu$g of $\lambda$DNA (purchased from New England Biolabs, 48kbp) was pelleted via ethanol precipitation, and subsequently resuspended in  120 $\mu$L of pH 7.5 buffer (20 mM Tris, 50 mM KCl, 2 mM $\mathrm{MgCl_2}$, 7$\%$ glycerol) yielding a final concentration (2.5 mg/mL) well within the entangled regime to emulate non-specific interactions under conditions analogous to \textit{in vivo}~\cite{fosado2023fluidification}. The solution was then incubated overnight at 37$^{o}$C to ensure complete dissolution of the DNA pellet. 

To study IHF-DNA interactions in acidic conditions and avoid DNA degradation and large pH fluctuations, we adjusted the pH immediately before the experiments by adding a pre-determined amount of diluted HCl to the concentrated $\lambda$-DNA solution. This method was carefully optimized through repeated buffer tests and serial HCl dilution. We found that using diluted HCl allowed larger volumes to be added ($\sim$ 10-12 $\mu$L), in turn minimizing the impact of pipetting inaccuracies and enabling more precise adjustment of the pH. The final pH was checked with a Mettler-Toledo pH meter and a micro-pH electrode before each experiment. 

\subsection{Time-resolved particle tracking microrheology}
For microrheology measurements, polystyrene microspheres (1.1$\mu$m, Sigma-Aldrich) were spiked into the solution of DNA with/without IHF. The microspheres were pelleted and resuspended in water adjusted to the target pH prior to adding them to the DNA solution. To ensure colloidal stability, microsphere suspensions at different pH values were examined by optical microscopy, confirming the absence of aggregation. Before use, the suspensions were sonicated for approximately 5 minutes to disrupt any transient clusters. The bead concentration was then adjusted such that each field of view contained approximately 20-30 particles, enabling consistent and reliable particle tracking during measurements. 

The glass slides were cleaned using Whatman lens-cleaning tissues (Cat No. 2105-841) with a few gentle wipes. Frame-seal incubation chambers (Bio-Rad, Cat. SLF0201) were adhered to the cleaned slides, and 5 $\mu$L of the sample was loaded into each chamber. Finally, the chambers were sealed with a 22 $\times$ 22 mm coverslip that had also been cleaned prior to use.

We perform recordings using Nikon T\textit{i}2 microscope with a 60x objective. Recordings of beads diffusing in the sample were started right after loading the solutions onto the glass slides. Movies were collected at 10 fps and spanning 3 minutes per position, and at least 5 positions per condition. To avoid
wall effects, we imaged at the mid-plane (100 $\mu$m) of the sample. 

We used TrackPy~\cite{crocker2000two} and custom-developed particle-tracking scripts (in Python and C++) to extract the trajectories of diffusing beads and compute the time-averaged mean-squared displacements (MSDs) as a function of lag time (t). Although the trajectories were recorded in 2D, the samples were isotropic; therefore, the displacements in the `x' and `y' directions were treated as independent random walks.

The diffusion coefficient $D$ is then computed by applying linear fits to the MSDs according to MSD $=$ (2d)Dt (with d=1, because the $x$ and $y$ directions are averaged together). From $D$ the viscosity of the fluid can be measured by using the Stokes-Einsten relation $\eta = k_B T/(3 \pi a D)$ with $a$ the diameter of the tracer. 

 \begin{figure*}[t!]
    \centering
    \includegraphics [width=1\textwidth] {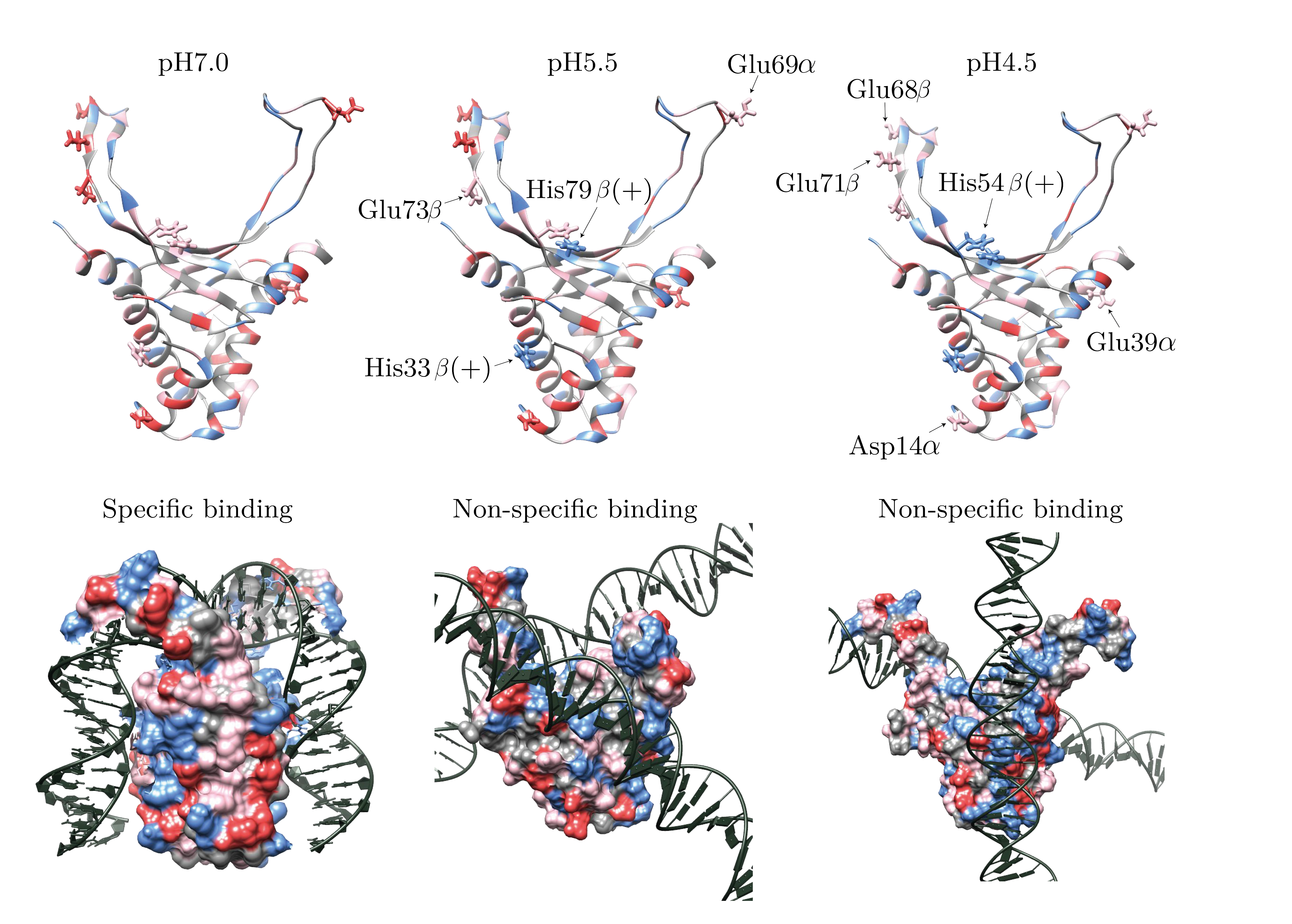}
    \caption{\textbf{All-atom simulations suggest pH-dependent protonation of IHF residues drives non-specific DNA binding.} \textbf{Top:} Close-up view of all the amino acid residues that change protonation state as pH decreases. Each residue is labeled with the pH value at which its protonation state changes relative to neutral pH. At lower pH, acidic residues (Asp and Glu) shift from negatively charged to neutral, while histidine (His) transitions from neutral to protonated.  \textbf{Bottom:} Comparison between the specific DNA–IHF complex (from PDB ID: 1IHF) and non-specific DNA bridging configurations obtained from all-atom MD simulations (this work). In all panels, DNA is shown in black, positively charged residues in blue, negatively charged in red, polar residues in pink, and apolar residues in gray.}
    \label{fig:allatom}
\end{figure*}

\subsection{All-atom simulations}

The Amber20 software package was employed for running molecular dynamics simulations via the CUDA implementation of the pmemd program~\cite{Amber20}. All simulations were solvated via TIP3P octahedral boxes with a minimum distance of 15{\AA}  between the solute and the box's edge. The systems were neutralized with a 0.15M concentration of KCl ions determined by the Dang parameters~\cite{dangions}. The ff14SB and parmBSC1 forcefields were employed to represent the protein and DNA, respectively~\cite{ff14SB,bsc1}. Simulations were performed under constant T and P (300 K and 1 atm) following standard protocols~\cite{24NanoscaleNoy}. The protonation states of ionizable residues at the two different initial structures (see below) were estimated at pH 4.5, 5.5, and 7.0 using the H++ web server, based on continuum electrostatics calculations within the Poisson-Boltzmann framework~\cite{H++}. 

To investigate unbiased binding events, the IHF structure (PDB 1IHF~\cite{rice1996crystal}) was positioned 2 nm away from a 50-bp DNA duplex with a random sequence (Figure S1). Five independent replicas were initiated from this configuration at each pH (15 replicas in total) and extended to 100 ns, following our previous approach~\cite{25JACSAuChan}. For simulations showing substantial protein-DNA interactions (i.e., more than one contact point), a second DNA duplex was placed 2 nm away, and the simulations were extended to 200 ns to evaluate bridge formation (Figure S1).

To test whether a bridging structure could be achieved from the specific binding mode, a second structure was generated by extracting the IHF-DNA complex bound at the H2 site (PDB 5J0N)~\cite{16eLIFEHJ} and retaining only the 11 bp at the center of the binding site. This complex was then inserted at the corresponding position of a linear 50-bp DNA to produce a structure in which the DNA remained unbent and was attached only to the extended $\beta$-ribbon arms of IHF, as done previously~\cite{Yoshua2021IHF,22CSBJWatson}. To prevent the transition to the specific DNA-bending complex, the inserted prolines were de-intercalated using a short series of umbrella-sampling simulations, and the DNA bases critical for intercalation were mutated. A 50-bp random DNA duplex was then positioned 2 nm away from this complex, and three 200-ns replicas were performed at each pH (Figure S1). Table~S1 provides an overview of all simulations, comprising a total of 24 runs with a cumulative simulation time of approximately 7~$\mu$s.

\section{Results and Discussion}

\subsection{All-Atom calculations suggest pH-dependent IHF protonation drives DNA bridging}

We performed electrostatics calculations to estimate which IHF residues could change their protonation state at different pH values. These calculations revealed that, as the pH decreased from 7.0 to 5.5, up to four amino acids became protonated, with an additional five residues protonating upon a further drop to pH 4.5 (with the specific residues depending on the conformation; see Methods and Figure~\ref{fig:allatom}). Because these nine residues are exposed on the protein surface, their protonation substantially reshapes the surface charge distribution.

 \begin{figure*}[t!]
    \centering
    \includegraphics [width=0.65\textwidth] {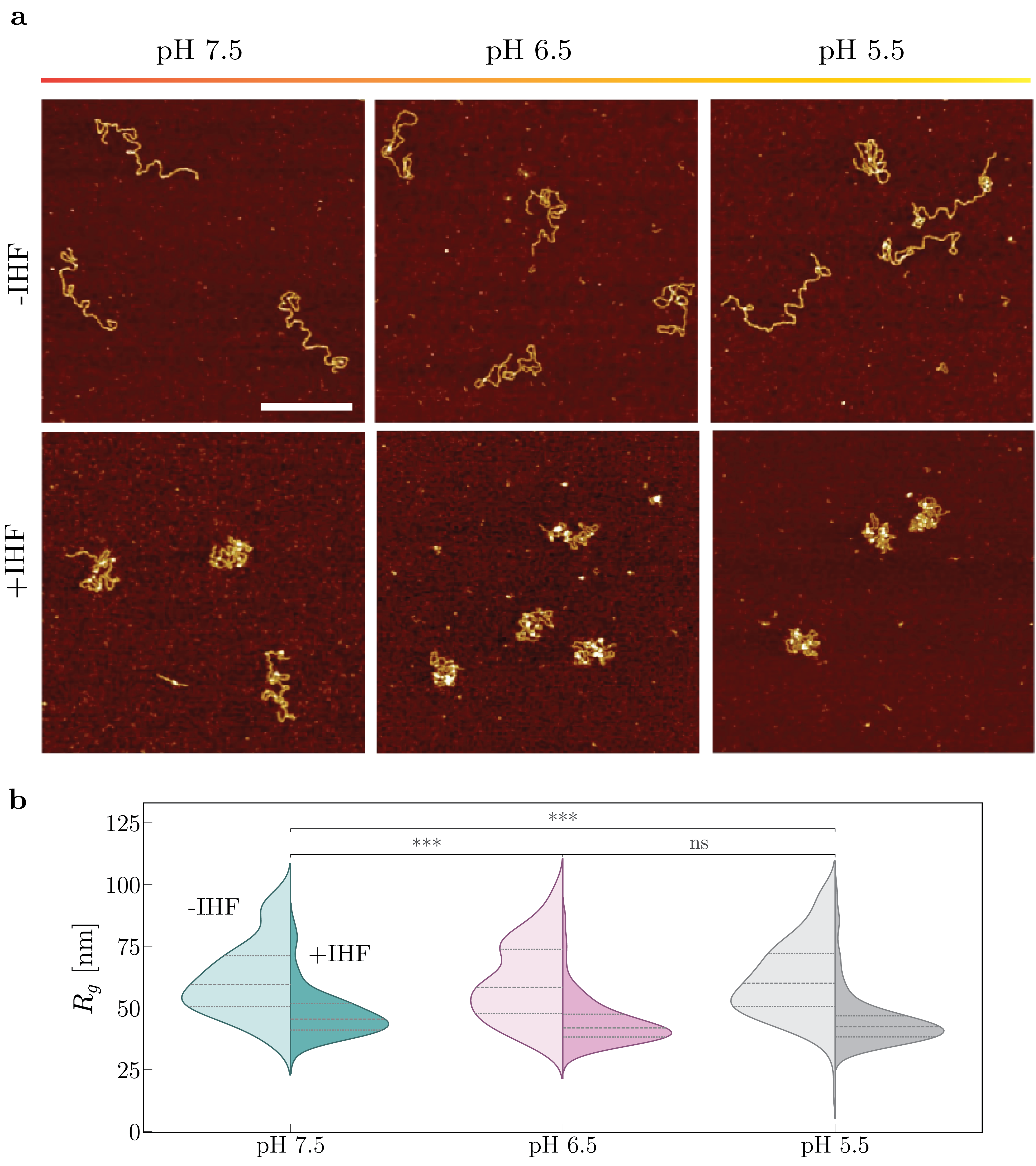}
    \caption{\textbf{Effect of pH on IHF-mediated DNA compaction.} \textbf{a.} AFM images of linearized pUC19 DNA at pH 7.5, 6.5, and 5.5 in the absence (top row) and presence (bottom row) of IHF. Scale bar corresponds to 100 nm \textbf{b.} Radius of gyration ($R_g$) across pH 7.5, 6.5, and 5.5, with Mann-Whitney U pairwise tests.
 $R_g$ distributions visualized as split violin plots (Left: Control/no IHF; Right: Sample with IHF). The mean $R_g$ was significantly smaller when IHF is present, and consistently small at lower pH for both, presence and absence, of IHF.
} 
    \label{radius_of_gyration}
\end{figure*}

To understand if these newly formed positively charged patches are sufficient to interact with negatively charged DNA we performed five independent replicas of all-atom simulations at each pH (4.5, 5.5, and 7.0). We initialised the system with the protein and DNA spatially separated, and observed spontaneous bridge formation (Movie S1-S4 and Figure S1). Strong bridges, spanning from the protein body to one $\beta$-ribbon arm, formed preferentially under acidic conditions (pH 4.5 and 5.5), whereas only a weak bridge was observed at pH 7.0 (see Figure~\ref{fig:allatom} and figS.MD.1stSet). 

We next tested whether bridging could also appear from a configuration mimicking the specific binding mode at neutral pH, as previously observed in simulations of supercoiled DNA~\cite{22CSBJWatson} (Figure~\ref{fig:allatom}). A second set of simulations was performed, with three replicas per pH, in which a second DNA duplex was introduced and placed separately from this IHF-DNA complex (Figure S1). Under these conditions, bridging was largely absent; the only exception was a single weak bridge at pH 7.0, involving just one contact point between the protein and the second DNA duplex (Figure S2). This indicates that this bridging mode is less probable compared with the mode observed in the first simulation set.

Overall, these results suggest that the protonation state of exposed residues in the IHF structure play a critical role in modulating its DNA-binding activity. These simulations provide insight into the underlying molecular mechanism, which appears to involve non-specific, electrostatic-driven interactions with the protein body. Such interactions are also reminiscent of the non-specific DNA-binding behavior reported for the closely related protein HU~\cite{16SciAdvAdhya,25JACSAuChan,remesh2020nucleoid}.

\subsection{AFM reveals IHF compacts DNA across all pH conditions}

To quantify how IHF modulates DNA conformation under different pH conditions, we performed AFM on linearized pUC19 in the presence/absence of IHF at pH 7.5, 6.5, and 5.5. Representative AFM micrographs are shown in Figure~\ref{radius_of_gyration}a. In the absence of IHF DNA molecules appeared as relaxed, extended filaments with random coil conformations characteristic of freely deposited linear DNA. Upon pre-incubation with IHF, all samples exhibited marked compaction, with molecules adopting bent or looped conformations, consistent with local DNA bending induced by IHF binding~\cite{rice1996crystal}. These structural changes were observed across all pH conditions, indicating that IHF retains strong DNA-binding activity even in acidic environments.

\begin{figure*}[t!]
    \centering
    \includegraphics [width=\textwidth] {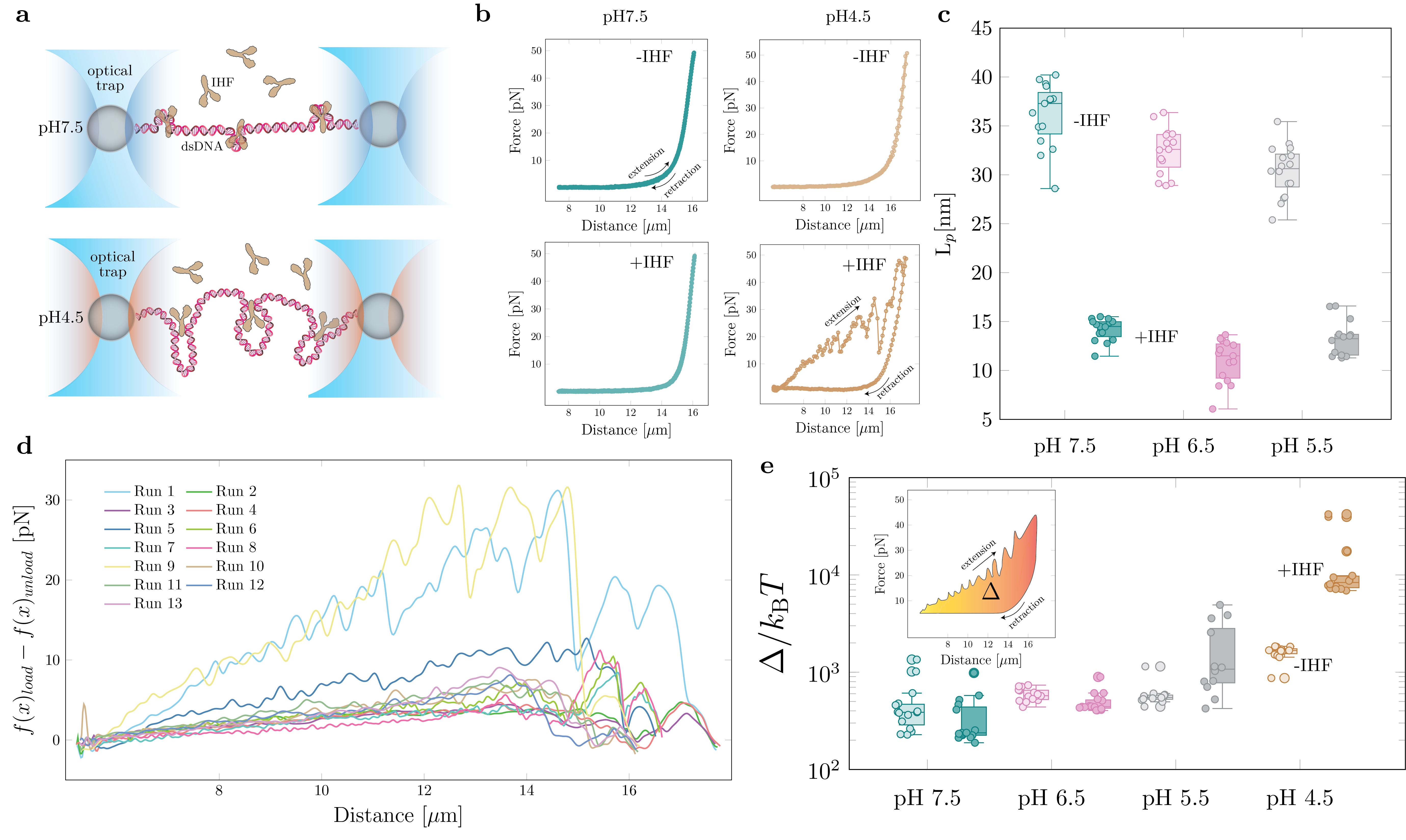}
    \caption{\textbf{DNA stretching reveals IHF-mediated DNA loops at low pH}. \textbf{a} Sketch of the optical tweezer experiment at neutral and low pH. \textbf{b} Force-extension curves recorded for $\lambda$-DNA incubated with and without IHF at different pH conditions. Note the presence of sawtooth pattern at low pH in presence of IHF, suggesting crosslinks and DNA loops that are broken in the extension phase. \textbf{c} Persistence length $L_p$ of $\lambda$DNA determined by fitting the curves to the worm-like chain model across conditions, showing effective softening (shorter $L_p$) due to IHF binding. \textbf{d} Curves obtained by subtracting the retraction (or ``unload'') curves to the extension (or ``load'') curves. Values larger than zero suggest hysteresis, i.e. that extension requires more force that the retraction. We define the area under this curve $\Delta$ and it represent the dissipated energy during the extension/retraction cycle. \textbf{e} Dissipated energy $\Delta$ across conditions and showing that at low pH (5.5 and 4.5) the presence of IHF creates significant DNA loops that need breaking to extend DNA.} 
    \label{Force_extension}
\end{figure*}

Quantitative analysis (see Methods) confirmed that IHF binding significantly reduced the radius of gyration of linearized pUC19 ($R_g$) relative to untreated controls (Figure~\ref{radius_of_gyration}b). For control DNA, $R_g$ values were distributed around $60$ nm at all pH levels [pH 7.5: $64.8 \pm 19.2$ nm; pH6.5: $62.2 \pm 19.9$ nm; pH5.5: $62.5 \pm 17.8$ nm], indicating similar molecular extension under neutral and mildly acidic conditions. In contrast, IHF-treated samples showed substantial compaction, with mean $R_g$ values decreasing to $47.7 \pm 10.0$ nm (26.4 $\%$ reduction) at pH7.5, $43.5 \pm 10.0$ nm (30.1 $\%$ reduction) at pH 6.5, and $44.6 \pm 13.1$ nm (28.6 $\%$ reduction) at pH5.5. The consistent downward shift in $R_g$ distributions indicates a robust compaction effect across the entire pH range tested.

At neutral pH, the observed conformations are consistent with the canonical IHF-induced minor-groove intercalation by its $\beta$-ribbon arms~\cite{rice1996crystal, dame2005role}. As pH decreases, the extent of compaction displays a mild increase, suggesting modestly enhanced binding affinity. This observation is in line with the protonation of acidic residues on the IHF surface, which increases the net positive charge and strengthens electrostatic attraction to the negatively charged DNA backbone. The presence of well-defined molecular contours, rather than amorphous aggregates, indicates that the observed compaction arises from organized IHF–DNA folding rather than nonspecific aggregation due to unstructured protein.

\subsection{IHF displays a pH-dependent transition from DNA bending to DNA bridging}

Building on the AFM observations, we next employed single-molecule optical tweezers assays to probe the mechanics of individual IHF-DNA complexes under tension. While AFM provides static ensemble snapshots, optical tweezers allow real-time measurement of the force-extension response of single $\lambda$-DNA tethers as IHF binding alters their elasticity.

Representative force-extension curves recorded in the absence and presence of IHF are shown in Figure~\ref{Force_extension}a-b. For naked $\lambda$-DNA (top panels), the force--extension relationship followed the canonical worm-like chain (WLC) model, displaying a smooth monotonic rise in force up to 50 pN as the molecule was extended from 8 $\mu$m to 16 $\mu$m. The persistence length ($L_p$) extracted from WLC fits was 36 nm, consistent with double-stranded DNA under physiological ionic conditions~\cite{smith1992direct}.

In the presence of IHF at neutral pH7.5, the force-extension curves shifted toward smaller extensions, and the fitted $L_p$ decreased by approximately 50\%, indicating increased flexibility due to protein-induced DNA bending. At pH 6.5, $L_p$ decreased further, consistent with enhanced binding affinity and local crowding along the DNA. This trend parallels the AFM results and suggests that under mildly acidic conditions, IHF populates DNA more densely, leading to cumulative bending and compaction.

\begin{figure*}[t!]
    \centering
    \includegraphics [width=0.65\textwidth] {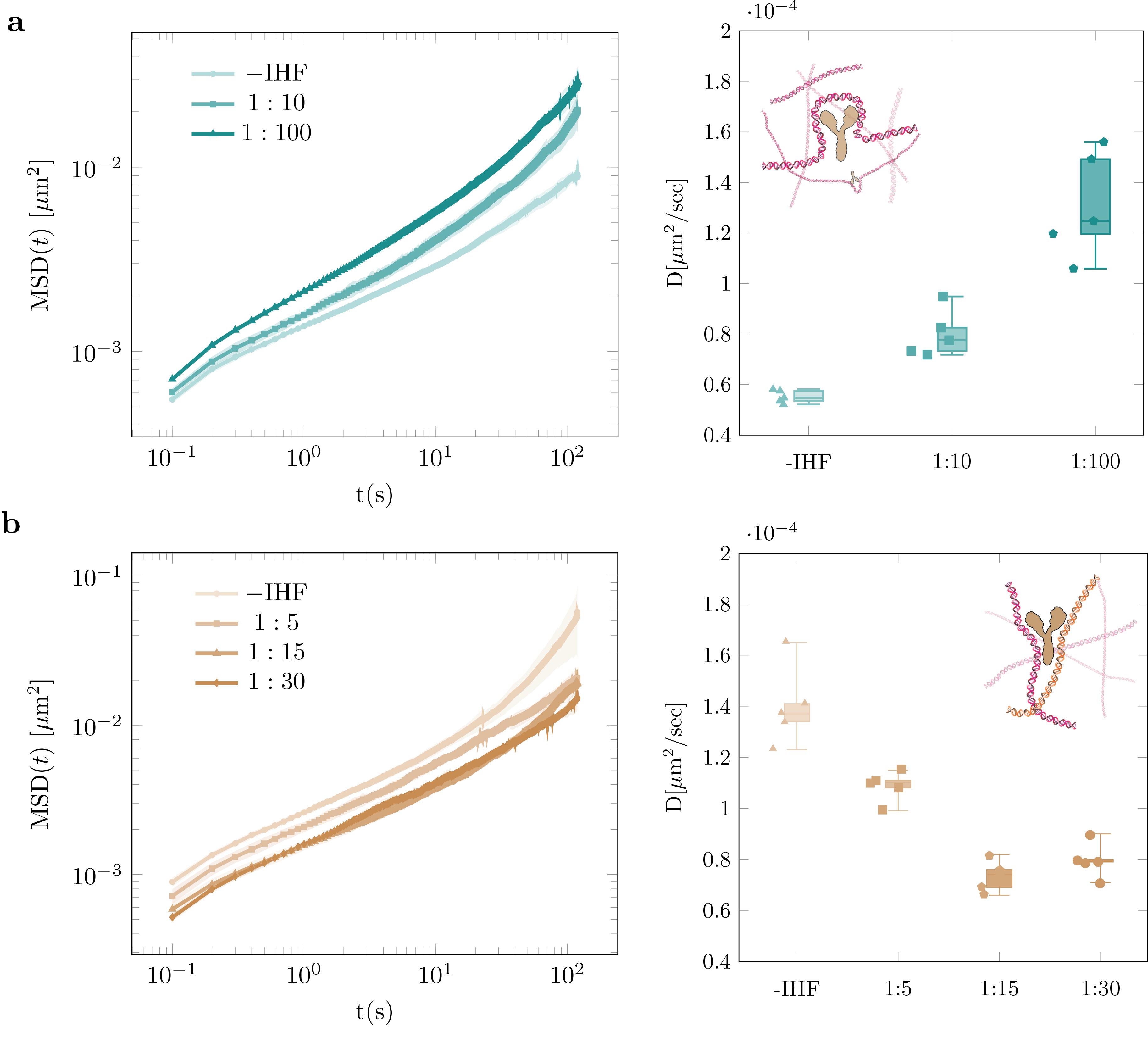}
    \caption{\textbf{Microrheology of IHF-DNA solutions} \textbf{a, left} Mean squared displacement of dense solutions of $\lambda$-DNA and IHF at neutral pH 7.5 and at different stoichiometries (DNA:protein ratio in legend).  \textbf{a, right} Diffusion coefficient extracted from the long time behaviour of the MSDs. \textbf{b, left} Mean squared displacement of dense solutions of $\lambda$-DNA and IHF at neutral pH 4.5 and at different stoichiometries (DNA:protein ratio in legend). \textbf{b, right} Diffusion coefficient extracted from the long time behaviour of the MSDs. These results sugges that IHF is a ``fluidifier'' at neutral pH and a ``thickener'' at low pH. } 
    \label{microrheo}
\end{figure*}

At pHs 5.5 and 4.5, a qualitatively distinct mechanical signature emerged: force-extension traces displayed irregular, sawtooth-like rupture events (Figure\ref{Force_extension}b, bottom right). These features were reproducible across replicates and absent in control DNA, signifying mechanical rupture of IHF-mediated DNA bridges. The magnitude and frequency of rupture events suggest multiple simultaneous binding interactions between separate DNA segments.

The area enclosed between loading and unloading curves ($\Delta$) was integrated to quantify dissipated mechanical work and normalized by $k_\mathrm{B}T$ to yield a dimensionless measure of energy. At neutral pH, $\Delta/k_\mathrm{B}T$ values were on the order of $10^{2}$--$10^{3}$, consistent with reversible elastic stretching of $\lambda$-DNA~\cite{cluzel1996dna}. Under acidic conditions (pH~5.5-4.5), $\Delta/k_\mathrm{B}T$ increased by more than an order of magnitude, reaching $\sim 10^4-10^5$, indicating that the mechanical work required to disrupt IHF-DNA contacts far exceeded thermal fluctuations. This confirms the formation of multivalent, load-bearing IHF bridges that resist force-induced dissociation.

The emergence of rupture peaks and elevated $\Delta/k_\mathrm{B}T$ at low pH reflects a pH-dependent switch in the mode of interaction. Protonation below the isoelectric point of the IHF $\alpha$ and $\beta$ subunits (pI~$\approx$~9.6~\cite{kozlowski2017proteome}) increases their positive charge, enhancing DNA attraction but also inducing steric and electrostatic crowding along the backbone~\cite{collette2023macromolecular}. Such conditions may promote cooperative clustering and bridging, generating networks that bear mechanical load until disrupted by high forces.

\subsection{The pH-dependent IHF switch leads to rheological thickening of DNA solutions}

To study the effects of the pH-switch observed in the previous section in crowded and entangled conditions of eDNA within biofilms, we performed passive microrheology on concentrated $\lambda$DNA-IHF mixtures at pH7.5 and pH4.5. This allowed us to quantify how IHF modulates the viscoelastic properties of entangled DNA networks as a function of pH. The trajectories of diffusing microspheres were tracked to obtain time-averaged mean-squared displacements 
\begin{equation}
    MSD(\tau) = \langle \left[ \bm{r}_i(t+\tau) - \bm{r}_i(t) \right]^2\rangle
\end{equation}
where $\bm{r}_i(t)$ is the position of tracer $i$ at time $t$ and the average is performed over initial times $t$ and tracers. From the MSD, we can extract the diffusion coefficient at long lag times as $D = \lim_{\tau \to \infty} MSD(\tau)/\tau$. Under neutral conditions (pH7.5; Figure~\ref{microrheo}a), the addition of IHF led to a speed up in the dynamics of the tracers and their long lagtime diffusion coefficient. 

This behaviour can be understood as a consequence of IHF-induced DNA bending, which effectively reduces the chain's persistence length and increases the entanglement length, $L_e$. As the number of entanglements per chain ($Z = L / L_e$) decreases, the network loses elastic constraints, leading to faster stress relaxation and enhanced tracer mobility. Molecular simulations and microrheology studies have shown this ``fluidification'' mechanism for IHF, where nanoscale DNA kinks increase $L_e$, promote polymer motion, and lower the elastic modulus of dense DNA solutions by orders of magnitude at physiological concentrations~\cite{fosado2023fluidification}. Thus, under near-neutral conditions, IHF acts as a DNA-bending protein that softens the DNA mesh through structural perturbation.

However, at acidic pH 4.5 (Figure~\ref{microrheo}b), the dynamics were reversed. The MSDs displayed a slowing down with addition of IHF, while $D$ decreased approximately two-fold at the higher protein:DNA ratio. This rheological thickening indicates that IHF forms more inter-molecular crosslinks at lower pH, consistently with the  optical tweezers observation. Interestingly, the DNA network appears to retain its fluid-like behaviour at large times (the tracers display freely diffusive behaviour at large time) thus suggesting that the IHF-mediated bridges are transient and are not sufficient to form a gel-like matrix.




\section{Conclusions}

This study uncovers a novel structural role for the abundant and ubiquitous nucleoid associated protein - IHF. Using a range of \textit{in vitro} and \textit{in silico} techniques, we have revealed that IHF switches from purely intramolecular DNA-bending to inter-molecular DNA-bridging at low pH, a condition typically found in biofilms.

More specifically, we used single molecule techniques to show that IHF induces DNA compaction across pHs (AFM, fig.~\ref{radius_of_gyration}) and that at low pH, protein-DNA bridges store energy disrupted during molecular pulling (optical tweezers, fig.~\ref{Force_extension}). Through modelling, we show that this switch can be attributed to the protonation of key exposed residues at low pH. All atom simulations confirm that protonation of those residues is sufficient to promote enhanced electrostatic interactions with the DNA phosphate backbone and to enable intermolecular cross-linking (Fig.~\ref{fig:allatom}).  

Given that IHF often works in entangled and crowded environments, we finally study the effect of low pH in mixtures of entangled DNA and IHF and again observe that in acidic conditions IHF acts as a ``thickener'', increasing the viscosity of the solution, rather than a ``fluidifier'', as observed at neutral pH (Fig.~\ref{microrheo}). 

This pH-dependent modulation of IHF-DNA interaction provides a mechanistic basis for the protein's dual role: as a chromatin organizer within bacterial nucleoids and as a structural component in biofilms. Our findings suggest that targeting IHF's bridging activity could destabilize biofilm matrices, particularly in cystic fibrosis pathogens~\cite{wilton2016extracellular}. Single-molecule techniques and biophysical quantification of IHF-DNA and IHF-bacterial biopolymer interactions could guide development of pH-optimized therapies for CF biofilms. Future work using non-consensus substrates like poly(dI-dC), combined with high-resolution structural and rheological measurements, may distinguish specific/nonspecific binding and support rational design of inhibitors.

To further understand the structural role of IHF in the mechanical stability of biofilms it will be interesting in the future to study the behaviour of IHF at low pH and in dense solutions of DNA mixed with other negatively charged biopolymers often found in biofilms. 

\section{Acknowledgements}
DM acknowledges the Royal Society and the European Research Council (grant agreement No 947918, TAP) for funding. The authors also acknowledge the contribution of EPSRC (EP/N027639/1 to A. N., EP/Y000501/1 to M.L. and EP/T022167/1 to J.L. and for computational resources), Generation Research (B.B) and 
of the COST Action Eutopia, CA17139.  We gratefully acknowledge support from the Microscopy core facility at the Wellcome centre at the University of Edinburgh funded through the Wellcome Discovery Research Platform for Hidden Cell Biology [226791]. For the purpose of open access, the author has applied a Creative Commons Attribution (CC BY) licence to any Author Accepted Manuscript version arising from this submission. 

\bibliographystyle{apsrev4-1}
\bibliography{journal}
\end{document}


\title{The Integration Host Factor is a pH-responsive protein that switches from DNA bending to DNA bridging in acidic conditions: Supplementary Information}

\author{Dinesh Parthasarathy}
\thanks{joint first author}
\affiliation{School of Physics and Astronomy, University of Edinburgh, Peter Guthrie Tait Road, Edinburgh EH9 3FD, United Kingdom}
\author{Saminathan Ramakrishnan}
\thanks{joint first author}
\affiliation{School of Physics and Astronomy, University of Edinburgh, Peter Guthrie Tait Road, Edinburgh EH9 3FD, United Kingdom}
\author{Georgia Tsang}
\affiliation{School of Biological Sciences, University of Edinburgh, Colin Maclaurin Road, Edinburgh EH9 3DW, United Kingdom}
\author{Auro Varat Patnaik}
\affiliation{School of Physics and Astronomy, University of Edinburgh, Peter Guthrie Tait Road, Edinburgh EH9 3FD, United Kingdom}
\author{Sabrina M.C. Hardy}
\affiliation{School of Physics and Astronomy, University of Edinburgh, Peter Guthrie Tait Road, Edinburgh EH9 3FD, United Kingdom}
\author{Jamieson Howard}
\affiliation{School of Physics, Engineering \& technology, University of York, YO10 5DD, United Kingdom}
\author{Willem Vanderlinden}
\affiliation{School of Physics and Astronomy, University of Edinburgh, Peter Guthrie Tait Road, Edinburgh EH9 3FD, United Kingdom}
\author{Braden Bylett}
\affiliation{School of Physics, Engineering \& technology, University of York, YO10 5DD, United Kingdom}
\author{James R.Law}
\affiliation{School of Physics, Engineering \& technology, University of York, YO10 5DD, United Kingdom}
\author{Mark C. Leake}
\affiliation{School of Physics, Engineering \& technology, University of York, YO10 5DD, United Kingdom}
\affiliation{Department of Biology, University of York, YO10 5DD, United Kingdom}
\author{Agnes Noy}
\thanks{corresponding author}
\affiliation{School of Physics, Engineering \& technology, University of York, YO10 5DD, United Kingdom}
\author{Davide Michieletto}
\thanks{corresponding author}
\affiliation{School of Physics and Astronomy, University of Edinburgh, Peter Guthrie Tait Road, Edinburgh EH9 3FD, United Kingdom}
\affiliation{MRC Human Genetics Unit, Institute of Genetics and Cancer, University of Edinburgh, Edinburgh EH4 2XU, United Kingdom}

\maketitle

\begin{figure*}[t!]
    \includegraphics[width=1\textwidth]{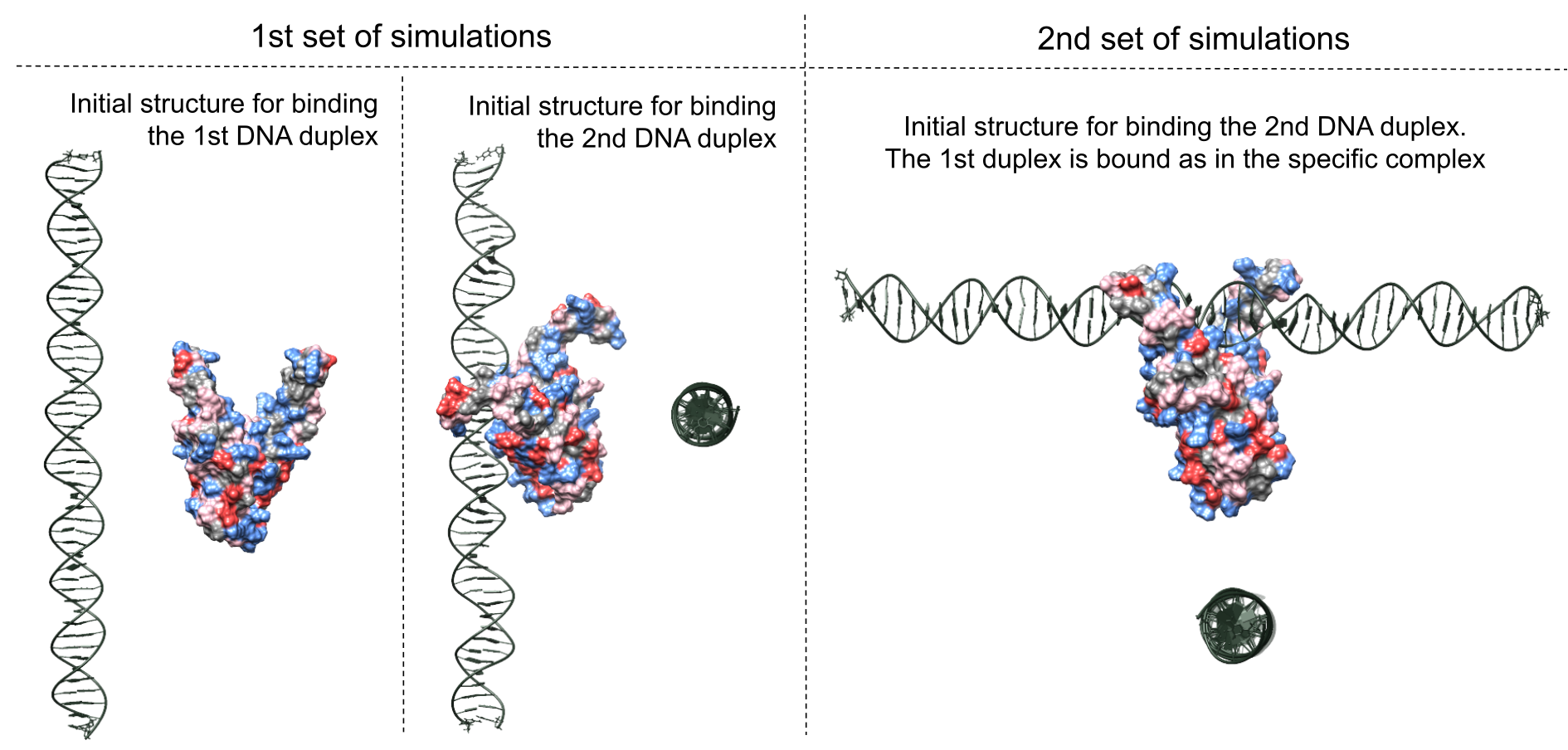}
    \caption{Initial structures for the two sets of simulations. In the first set (left), DNA bridging by IHF occurs in two steps: the protein is placed 2 nm from a DNA duplex, and if an interaction forms, the resulting complex is positioned 2 nm from a second duplex. In the second set (right), the protein already interacts with the first duplex as in the specific binding complex, but without high-affinity interactions (DNA unbent, prolines not intercalated). A second DNA duplex is then placed nearby to attempt bridging. In both sets, the second DNA duplex is positioned perpendicular to the first duplex to minimize electrostatic repulsion. DNA is shown in black, positively charged residues in blue, negatively charged in red, polar residues in pink, and apolar residues in gray.}
    \label{figS:MD_IS}
\end{figure*}

\begin{figure*}[t!]
    \includegraphics[width=1\textwidth]{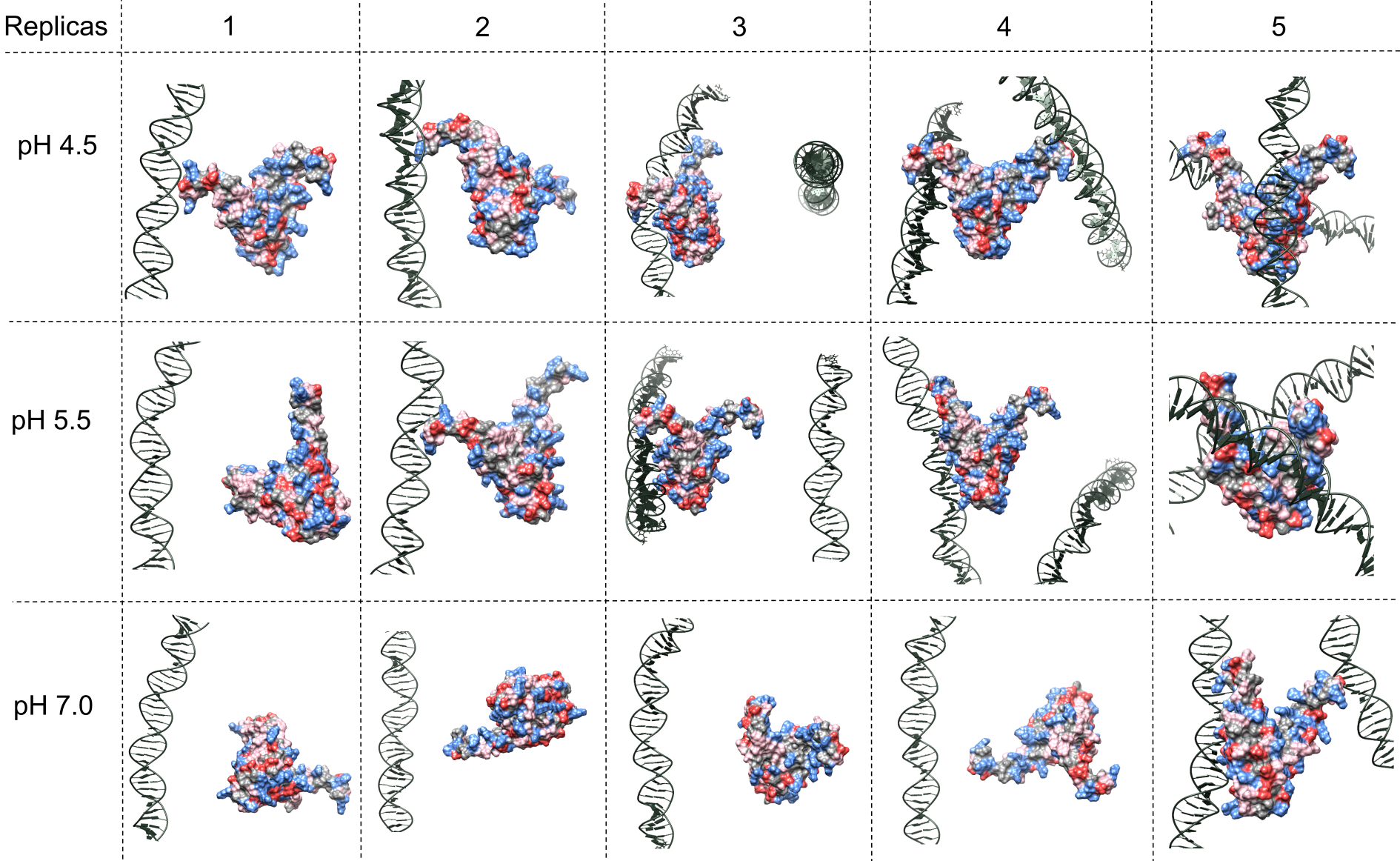}
    \caption{Final frames from the 15 simulations of the first set (5 replicates per pH). When the first DNA duplex forms strong interactions with the protein (i.e., multiple contact points), a second DNA duplex is introduced to evaluate additional binding. Strong bridging is observed only in the last replicas at pH 4.5 and 5.5; these frames are also shown in Figure 2. DNA is shown in black, positively charged residues in blue, negatively charged in red, polar residues in pink, and apolar residues in gray.}
    \label{figS:MD_1stS}
\end{figure*}

\begin{figure*}[t!]
    \includegraphics[width=0.8\textwidth]{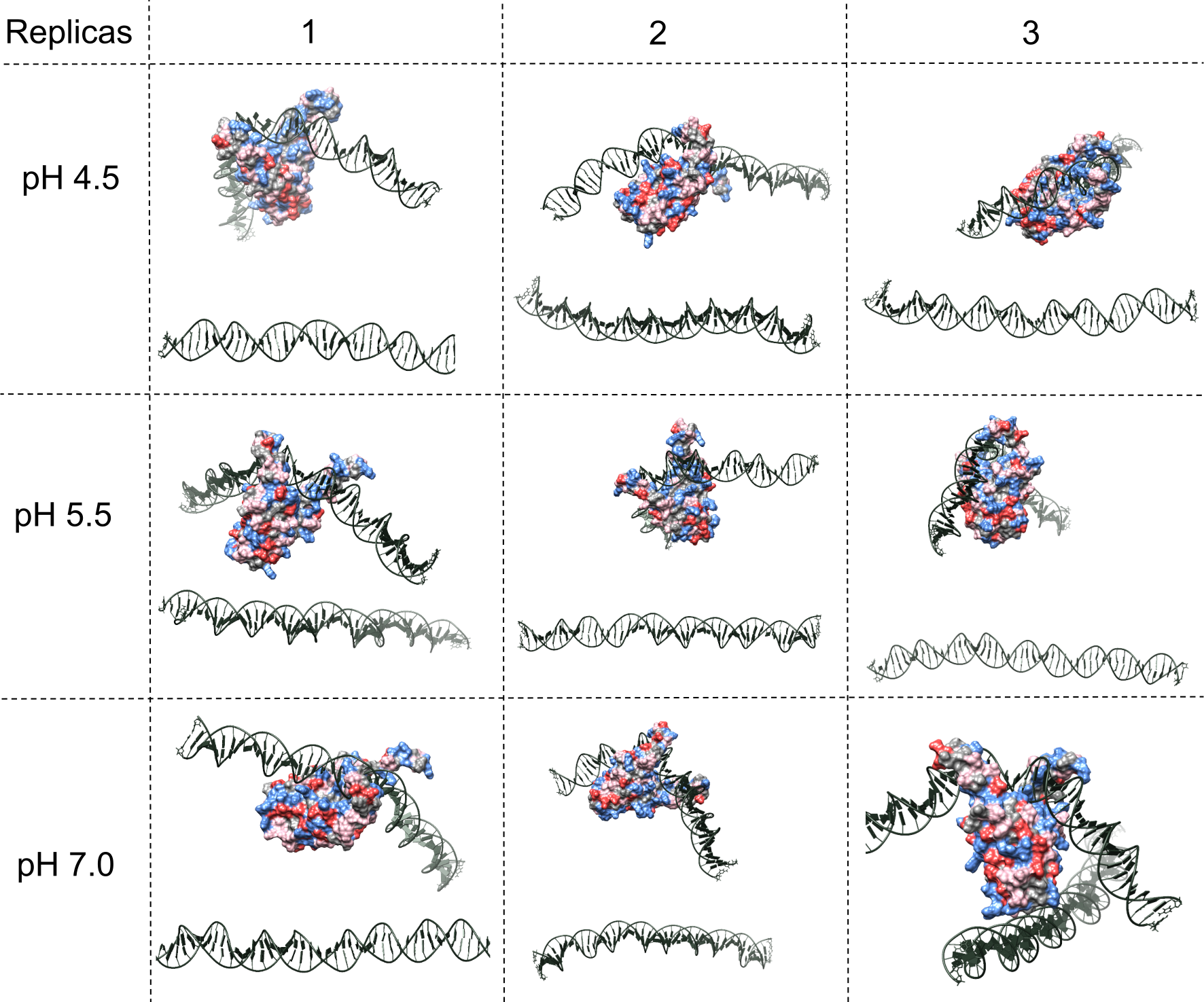}
    \caption{Final frames from the 9 simulations of the second set (three replicates per pH). Only a single, weak bridging event is observed. DNA is shown in black, positively charged residues in blue, negatively charged in red, polar residues in pink, and apolar residues in gray.}
    \label{figS:MD_2ndS}
\end{figure*}

\begin{table*}[t!]
    \begin{tabular}{|c|c|c|c|}
        \hline
		\; pH \; & \; Replica \; & \; Binding to 1st DNA duplex \; & \; Binding to 2nd DNA duplex \; \\
		\hline
        \multicolumn{4}{|c|}{1st set of simulations} \\
        \hline
        \multirow{4.5} & 1 & Weak (one $\beta$-ribbon arm) & - \\ 
        & 2 & Weak (one $\beta$-ribbon arm) & - \\ 
        & 3 & \; Strong (protein body and arm) \; & No interaction \\
        & 4 & Strong (protein body and arm) & Weak (one $\beta$-ribbon arm) \\
        & 5 & Strong (protein body and arm) & \; Strong (protein body and arm) \; \\
        \hline
        \multirow{5.5} & 1 & No interaction & - \\ 
        & 2 & Weak (one $\beta$-ribbon arm) & - \\ 
        & 3 & Strong (protein body and arm) & No interaction \\
        & 4 & Strong (protein body and arm) & No interaction\\
        & 5 & Strong (protein body and arm) & Strong (protein body and arm) \\
        \hline
        \multirow{7.0} & 1 & No interaction & - \\ 
        & 2 & No interaction & - \\ 
        & 3 & No interaction & - \\
        & 4 & No interaction & - \\
        & 5 & Strong (protein body and arm) & Weak (one $\beta$-ribbon arm) \\
        \hline
        \multicolumn{4}{|c|}{2nd set of simulations} \\
        \hline
        \multirow{4.5} & 1 & - & No interaction \\ 
        & 2 & - & No interaction \\ 
        & 3 & - & No interaction \\
        \hline
        \multirow{5.5} & 1 & - & No interaction \\ 
        & 2 & - & No interaction \\ 
        & 3 & - & No interaction \\
        \hline
        \multirow{7.0} & 1 & - & No interaction \\ 
        & 2 & - & No interaction \\ 
        & 3 & - & Weak (protein's 'bottom') \\
        \hline
    \end{tabular}
    \caption{Summary of all simulations performed at atomic resolution. Representative structures from each are shown in Figures S\ref{figS:MD_1stS} and S\ref{figS:MD_2ndS}. Protein-DNA interactions were classified by visual inspection as strong (multiple contact points) or weak (single contact point), with the interacting protein region indicated in parentheses. A second DNA duplex was introduced to evaluate additional binding only when the first duplex formed a strong interaction with the protein.}
\end{table*}

Movie S1. First stage of the bridging process for the fifth replica at pH 4.5 in the first set of simulations, where a DNA duplex interacts non-specifically with IHF (see Table S1).\\

Movie S2. Second stage of the bridging process for the fifth replica at pH 4.5 in the first set of simulations, where a second DNA duplex interacts non-specifically with IHF, resulting in the formation of a DNA-protein-DNA bridge.\\

Movie S3. First stage of the bridging process for the fifth replica at pH 5.5 in the first set of simulations, where a DNA duplex interacts non-specifically with IHF (see Table S1).\\

Movie S4. Second stage of the bridging process for the fifth replica at pH 5.5 in the first set of simulations, where a second DNA duplex interacts non-specifically with IHF, resulting in the formation of a DNA-protein-DNA bridge.\\

\bibliographystyle{apsrev4-1}
\bibliography{journal}